\documentclass{article}
\usepackage{float}
\usepackage{flafter}

\usepackage[margin=1in]{geometry}

\usepackage[dvips]{epsfig}
\usepackage{mathtools}
\usepackage{natbib}
\usepackage{xcolor}
\usepackage{dsfont}
\usepackage{tikz}
\usepackage{pict2e}

\usepackage{amsmath}
\usepackage[utf8]{inputenc}
\usepackage[english]{babel}
\usepackage{graphicx}
\usepackage{comment}
\usepackage{epic}
\usepackage{lscape}
\usepackage{rotating}
\usepackage{epstopdf}
\usepackage{wrapfig}
\usepackage{amssymb}
\usepackage{bm}

\usepackage{booktabs}

\usepackage{array}

\newcommand\lrr[1]{\ensuremath{		\left( #1 \right)	}}

\usetikzlibrary{math} 
\usetikzlibrary{shapes.misc}
\tikzset{cross/.style={cross out, draw=black, minimum size=2*(#1-\pgflinewidth), inner sep=0pt, outer sep=0pt},
	cross/.default={3pt}}

\renewcommand{\geq}{\geqslant}
\renewcommand{\leq}{\leqslant}

\usepackage{overpic}

\usepackage[normalem]{ulem}

\usepackage{soul}

\title{Stokes waves in rotational flows: \\
internal stagnation and overhanging profiles}
\author
{A.~Doak$^1$
\thanks{Email address for correspondence: add49@bath.ac.uk},
V.~M.~Hur$^2$,
and J.-M.~Vanden-Broeck$^3$}
\date{%
    \small {$^1$Department of Mathematical Sciences, University of Bath, Bath, BA2 7AY, UK}\\ %
    $^2$Department of Mathematics, University of Illinois Urbana-Champaign, Urbana, IL, 61801, USA \\ 
    $^3$Department of Mathematics, University College London, London, WC1E 6BT, UK \\ 
    \today
}

\begin{document}

\maketitle

\begin{abstract}
Periodic travelling waves at the free surface of an incompressible inviscid fluid in two dimensions under gravity are numerically computed for an arbitrary vorticity distribution. The fluid domain over one period is conformally mapped from a fixed rectangular one, where the governing equations along with the conformal mapping are solved using a finite difference scheme. This approach accommodates internal stagnation points, critical layers, and overhanging profiles, thereby overcoming limitations of previous studies. The numerical method is validated through comparisons with known solutions for zero and constant vorticity. Novel solutions are presented for affine vorticity functions and a two-layer constant vorticity scenario.
\end{abstract}

\section{Introduction}\label{sec:intro}

Much of the theory of free-surface water waves assumes that the flow is irrotational, an assumption well justified in some applications. Also, in the absence of initial vorticity, boundaries, or external forces, vorticity will remain zero for all times. Excellent surveys of travelling water waves can be found in \cite{Toland1996}, \cite{StraussBAMS2010}, \cite{vanden2010gravity}, and \cite{QAM2022}. However, rotational effects play a crucial role in numerous physical situations. For example, in any region where wind is blowing, there is a surface drift of the water and wave characteristics such as maximum wave height are sensitive to the velocity in the wind-drift boundary layer. See, for instance, \cite{da1988steep} for further discussion.

A special case of free-surface waves with vorticity was found by \citet{gerstner1802, gerstner1809theorie}, who derived an exact solution for periodic travelling waves in infinite depth for some nonzero vorticity. Notably, no such closed-form solution exists for zero vorticity. Many authors have also considered the case of constant vorticity. Among the most striking in constant vorticity flows are profiles with multi-valued height and even profiles which intersect themselves tangentially above the trough, enclosing an air bubble. By contrast, in an irrotational flow, the wave profile must be the graph of a single-valued function  (see \citeauthor{HW2022JDE} \citeyear{HW2022JDE}, and references therein).

Nonzero vorticity introduces substantial challenges in the mathematical treatment of travelling water waves. The stream function is no longer harmonic, whence complex analysis techniques are not directly applicable. Also, the fluid surface is a priori unknown, and like in all free boundary problems, one must fix the free boundary. \citet{dubreil1934determination} proposed a semi-hodograph transformation, which interchanges the vertical coordinate (independent variable) with the stream function (dependent variable), successfully proving the existence of small-amplitude solutions. \citet{constantin2004exact} took the matters further and established the existence of large-amplitude solutions. 
The semi-hodograph transformation has the advantage that it accommodates arbitrary vorticity distributions. Its disadvantage is that it cannot permit internal stagnation points, critical layers or overhanging profiles. Nonetheless, it has been widely employed in analytical studies \citep[see, for instance,][and references therein]{StraussBAMS2010} and numerical investigations \citep{dalrymple1977numerical,thomas1990wave,ko2008large,ko2008effect,amann2018numerical}.
Another approach is the so-called flattening transformation, for which the vertical coordinate is scaled to a fixed height. It can successfully describe interior stagnation points and critical layers, but it remains incapable of treating overhanging profiles.

A great deal of research on travellling water waves focuses on constant vorticity. This is because of its analytical tractability: The fluid domain can be conformally mapped from a fixed one, where the governing equations can be formulated in terms of quantities at the fluid surface, similarly to zero vorticity. Numerical studies \citep{simmen1985steady,da1988steep,pullin1988finite,vanden1994solitary, vanden1996periodic,DH1,dyachenko2019stokes, ribeiro2017flow, guan2020particle} and rigorous analysis \citep{CSV2016, KKL2020, hur2020exact, HW2022JDE} \citep[see also][and references therein]{QAM2022}, employing a conformal transformation, have revealed remarkable phenomena, including overhanging profiles, interior stagnation points, and critical layers. These features are expected to persist for more general vorticity distributions, motivating the development of numerical methods capable of capturing such behaviors. 

Here we employ a conformal transformation introduced and analytically examined by \citet{wahlen2022large, wahlen2023global}, which accommodates interior stagnation points, critical layers, and overhanging profiles for arbitrary vorticity distributions. Unlike \cite{wahlen2022large}, who reduced the problem to a nonlocal equation, we numerically solve (local) partial differential equations in a fixed rectangular domain using a finite difference scheme. The accuracy of our numerical method is validated through comparisons with known solutions for zero and constant vorticity, where strong agreement is observed. Additionally, novel solutions are present for affine vorticity functions and a two-layer constant voticity scenario. 

To the best of the authors' knowledge, this is the first numerical study of Stokes waves in an arbitrary vorticity flow, capturing the effects of critical layers and overhanging profiles. The solutions obtained herein can provide a basis for stability analysis and other applications. While the present approach accounts only for gravitational effects, it can be extended to incorporate additional physical effects such as density stratification, surface tension, and hydroelasticity. 

\section{Formulation}\label{sec:formulation}

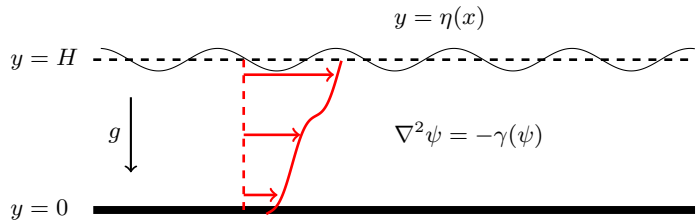
\begin{figure}
	\centering
	\begin{tikzpicture}
		\draw [ -,line width=3pt] (0,0) -- (8,0) ;
		\draw [ -,line width=1pt] (0,2) -- (8,2)[dashed] ;
		\draw[thin]  plot [samples=300,domain=0.1:8] (\x,{2+.3*sin(2*(2*\x +pi/2-pi-.95) r)/2});
		\node[text width=2cm] at (-.1,2) {\small{$y=H$}};
		\node[text width=2cm] at (-.1,0) {\small{$y=0$}};
		\node[text width=4cm] at (6,1) {\small{$\nabla^2 \psi=-\gamma(\psi)$}};
		\node [text width=6cm] at (7,2.55) {\small{$y=\eta(x)$}};
		\draw [ ->,thick]  (.5,1.5) -- (.5,.5)  node [midway,left] {\small{$g$}};
		\draw[ red,-,line width=1pt]  plot [samples=300,domain=3.3:4.3] (\x-1,{.3*sin(2*(5*\x +pi/2-pi-.95) r)/2   +  2.05*(\x-3.3) - .1});
		\draw [ red,-,line width=1pt] (2,0) -- (2,2) [dashed] ;
		\draw [ red,->,line width=1pt] (2,0.2) -- (2.45,.2) ;
		\draw [ red,->,line width=1pt] (2,1) -- (2.75,1) ;
		\draw [ red,->,line width=1pt] (2,1.8) -- (3.2,1.8) ;
	\end{tikzpicture}	
	\caption{Schematic of the fluid domain in a reference frame moving at the wave speed, where the free surface is given by $y=\eta(x)$ and the effects of nontrivial vorticity are represented by the background shear flow, illustrated with horizontal arrows.}\label{fig:conf_z}
\end{figure}

We consider periodic travelling waves at the free surface of an incompressible inviscid fluid in two dimensions under the influence of gravity in a rotational flow. Suppose that in Cartesian coordinates, the $x$-axis points in the direction of wave propagation and the $y$-axis opposite to gravity. In a frame of reference moving with the wave, the fluid flow appears steady and occupies a domain bounded above by a free boundary $y=\eta(x)$ and below by a fixed boundary $y=0$. Let 
\[
\Omega_{(x,y)}=\{(x,y)\in\mathbb{R}^2: -L/2\leq x\leq L/2, 0<y<\eta(x)\},
\]
where $L>0$ is the wavelength, and 
\[
H=\int_{-L/2}^{L/2} \eta(x)~dx
\]
the mean fluid depth. We introduce a stream function $\psi(x,y)$ such that the velocity of the fluid is given by $(\psi_y,-\psi_x)$, where subscripts denote partial differentiation. The vorticity satisfies
\[
\text{(vorticity)}=-\nabla_{(x,y)}^2\psi(x,y)=\gamma(\psi(x,y)), 
\]
where $\gamma$ is an arbitrary function of one variable. Throughout we assume it is single-valued. The relative mass flux is defined as
\[
Q=\int^{\eta(x)}_0 \psi_y(x,y)~dy,
\]
which is independent of $x$. The Stokes wave problem then takes the form:
\begin{subequations}\label{eq:(x,y)}
\begin{align}
&\nabla_{(x,y)}^2\psi=-\gamma(\psi)&& \text{in $\Omega_{(x,y)}$}, \label{eq:feqz}\\
&|\nabla_{(x,y)}\psi|^2+2g\eta=B  && \text{at $y=\eta(x)$}, \label{eq:dbcz}\\
&\psi=Q && \text{at $y=\eta(x)$}, \label{eq:kbcz} \\
&\psi=0 && \text{at $y=0$}, \label{eq:kbcz2}
\end{align}
\end{subequations}
where $g$ is the constant of gravitational acceleration and $B$ the Bernoulli constant. Throughout we assume the wave profile is symmetric about $x=0$. See Figure~\ref{fig:conf_z}. 

While \eqref{eq:(x,y)} accounts only for gravitational effects, it can be easily extended to include additional physical effects such as capillarity and hydroelasticity.

\begin{figure}
\centering
\begin{tikzpicture}[domain=-2*pi:2*pi]
			\tikzmath{\Lb=8; \Hb=2; \gap=1.7*\Lb; \M=10; \N=8; \Mm=\M-1; \Mmm=\M-2; \Nm=\N-1; \Nmm=\N-2; \Dx=\Lb/\M; \Dy=\Hb/\N; \bx=.4*\Lb/\M; \by=.4*\Hb/\N;}
			\node (A) at (0,0) {};
			\node (B) at (\Lb,0) {};
			\node (C) at (0,\Hb) {};
			\node (D) at (\Lb,\Hb) {};	
			\draw[draw=black] (0,0) rectangle ++(\Lb,\Hb);
			\node at (0.5*\Lb,0.6*\Hb) {$\nabla^2 Y = 0$};
			\node at (0.5*\Lb,-0.25) {$Y = 0$,  \quad $\psi=0$};
			\node at (0.5*\Lb,\Hb+0.25) {$\frac{1}{J}  \psi_\beta^2 + 2 gY= B$, \quad $\psi=Q$};

			\node at (0.5*\Lb,0.4*\Hb)  {$\nabla^2 \psi = -J \gamma(\psi)$};
			\node [rotate=90] at (+0.75,0.5*\Hb) {$\psi_\alpha = 0 $};
			\node [rotate=90] at (\Lb-0.75,0.5*\Hb) {$\psi_\alpha = 0 $};
			\node [rotate=90] at (+0.3,0.5*\Hb) {$Y_\alpha = 0$};
			\node [rotate=90] at (\Lb-0.3,0.5*\Hb) {$Y_\alpha = 0 $};
			\node at (0,-0.25) {$\alpha=-\pi$};
			\node at (\Lb,-0.25) {$\alpha=\pi$};
			\node at (-.75,0) {$\beta=-d$};
			\node at (-.75,\Hb) {$\beta=0$};
\end{tikzpicture}	
\caption{The conformal parametriasation of the fluid domain in the rectangular region $\Omega_{(\alpha,i\beta)}$. The formulation consists of two field equations and two sets of boundary conditions for the two unknowns $\psi$ and $Y$. 
}\label{fig:formab}
\end{figure}

In what follows, we identify $\mathbb{R}^2$ with $\mathbb{C}$ whenever convenient. 
Following \citep{wahlen2022large, wahlen2023global}, we introduce 
\begin{equation}\label{eq:conformal}
x=X(\alpha,\beta)\quad \text{and}\quad y=Y(\alpha,\beta),
\end{equation}
which conformally maps 
\[
\Omega_{(\alpha,\beta)}=\{\alpha+i\beta\in\mathbb{C}: -\pi\leq \alpha\leq \pi, -d<\beta<0 \}
\]
for some $d>0$, the conformal depth, to the fluid domain $\Omega_{(x,y)}$ in the physical variables, satisfying $\nabla_{(\alpha,\beta)}^2 X, \nabla_{(\alpha,\beta)}^2 Y=0$. 
Additionally, they satisfy the Cauchy--Riemann equations, $X_\alpha = Y_\beta$ and $X_\beta=-Y_\alpha$. A straightforward calculation reveals
\[
\psi_x = \frac{1}{J}(\psi_\alpha Y_\beta - \psi_\beta Y_\alpha )
\quad\text{and}\quad 
\psi_y = \frac{1}{J}(\psi_\alpha Y_\alpha + \psi_\beta Y_\beta),
\]
where $J= |\nabla_{(\alpha,\beta)} Y|^2$, whence $\nabla_{(\alpha,\beta)}^2 \psi = J \nabla_{(x,y)}^2 \psi$. Furthermore, 
\[
|\nabla_{(x,y)}\psi|^2 
= \frac{1}{J}  \psi_\beta^2 \quad \text{at $\beta=0$}.
\]
Substituting these into \eqref{eq:(x,y)}, and imposing periodicity in the $\alpha$-variable, we arrive at:
\begin{subequations}\label{eq:psi}
\begin{align}
&\nabla^2 \psi = -J \gamma(\psi)  && \text{in } \Omega_{\xi}, \label{eq:feq1}\\
&\psi=Q && \text{at } \beta=0, \label{eq:dirichlet1} \\
&\psi=0 && \text{at } \beta=-d, \label{eq:dirichlet2}  \\
&\psi_\alpha= 0 && \text{at } \alpha=\pm\pi, \label{eq:neumann1} 
\end{align}
\end{subequations}
and
\begin{subequations}\label{eq:Y}
\begin{align}
&\nabla^2 Y = 0 && \text{in } \Omega_{\xi}, \label{eq:feq2}\\
&\frac{1}{J}  \psi_\beta^2 + 2 gY = B && \text{at } \beta=0, \label{eq:dbc}\\
&Y=0 && \text{at } \beta=-d, \label{eq:dirichlet3}  \\
&Y_\alpha= 0 && \text{at } \alpha=\pm\pi, \label{eq:neumann2} 
\end{align}
\end{subequations}
where $\nabla^2=\nabla^2_{(\alpha,\beta)}$.
See Figure~\ref{fig:formab}. 

If both the wavelength $L$ and the mean fluid depth $H$ are fixed, the relative mass flux $Q$ and the conformal depth $d$ must be allowed to vary. If only one of $L$ or $H$ is prescribed, on the other hand, $d$ can remain fixed while $Q$ varies. See Section~\ref{sec:method} for further discussion of bifurcation parameters.

Unlike the semi-hodograph transformation \citep{dubreil1934determination, constantin2004exact}, the conformal mapping does not require streamlines to be graph-like in the $x$-variable, thereby allowing for overhanging profiles. Also, internal stagnation points do not result in singularities in the mapping and therefore are admissible. However, care must be taken when stagnation points occur at the boundary of the fluid domain. 

An important limiting scenario for travelling water waves involves the formation of a stagnation point at the fluid surface, characterised by a  $120^\circ$ angle at the crest. Since \eqref{eq:conformal} is conformal, singularities cannot arise in the interior of the fluid domain, but they may still develop at the boundary. Specifically, mapping a $120^\circ$ angle in $z:=x+iy$ to a straight line in $\zeta:=\alpha+i\beta$ would exhibit $z\sim \zeta^{3/2}$ to leading order in the vicinity of a stagnation point at the fluid surface and, hence, the second-order derivatives $Y_{\alpha\alpha}$ and $Y_{\beta\beta}$ become singular near the stagnation point. 
Consequently, our numerical method fails to converge for near-limiting and limiting solutions with an angle at the wave crest. See Section~\ref{sec:results} for further discussion of numerical limitations. An interesting direction for future research is to improve our numerical strategy to effectively resolve such singular behaviour. 

It is interesting how our formulation simplifies under some assumptions. When the flow is irrotational, \eqref{eq:psi} is trivially satisfied by $\psi=\frac{Q}{d}\beta$ because a unique conformal mapping from the fluid domain to a fixed rectangular region is a complex potential $\phi+i\psi$, where $\phi$ is a velocity potential such that the velocity $=(\phi_x,\phi_y)$. Therefore, solving \eqref{eq:Y} suffices to determine the solution in the absence of vorticity. 

For a rigid wall, on the other hand, \eqref{eq:dbc} is replaced by $Y=H$ and \eqref{eq:Y} is trivially satisfied by $Y=H+\frac{H}{d}\beta$. Therefore, it remains to solve the vorticity equation. This corresponds to waves propagating in a shear flow in a uniform fluid channel. A promising extension of our numerical approach is to incorporate baroclinic vorticity arising from density stratification, which would be to include the free-surface effects in internal wave models. 
See, for instance, \citep{stastna2022internal} and references therein.

\section{Linear theory}\label{sec:linear}

Before addressing nonlinear solutions, we examine the linear theory, which is relevant to small-amplitude perturbations of a flat surface. The dispersion relation for free-surface water waves in non-constant vorticity flows has been investigated in several studies, including \cite{ehrnstrom2011steady}, \cite{karageorgis2012dispersion}, \and 
\cite{kozlov2014steady}. Here we present a summary of the theory for completeness and outline a numerical solution method. It is important to note that the present analysis concerns steady waves and does not address stability, which would require working with the time-dependent problem where the vorticity function cannot be prescribed. The stability of Stokes waves in non-constant vorticity flows has been previous studied, for instance, in \cite{hur2008unstable}.  

Let 
\[
\psi(x,y) = \psi_0(y) + \epsilon \psi_1(x,y)+\cdots
\quad\text{and}\quad \eta(x) = \epsilon \eta_1(x)+\cdots
\]
for $|\epsilon|\ll 1$, where $\psi_0(y)$ represents the background shear flow and is independent of $x$. Substituting these into \eqref{eq:(x,y)} and collecting terms at leading order, we gather 
\begin{equation}\label{eq:psi0}
\left\{
\begin{aligned}
&\frac{d^2\psi_0}{dy^2} = -\gamma(\psi_0) \quad \text{for $0<y<H$} \\
&\psi_0(H) = Q\quad\text{and}\quad \psi_0(0) = 0. 
\end{aligned}\right.
\end{equation}
For some choices of $\gamma(\psi)$, an analytical solution is possible. In general, however,  a numerical approach, such as using a shooting method, is required. At the order of $\epsilon$, we gather
\[
\left\{\begin{aligned}
&\nabla^2 \psi_1=  -\gamma'(\psi_0)\psi_1 && \text{for $0<y<H$}  \\
&\psi_0'(H) {\psi_1}_y + \psi_0'(H) \psi_0''(H) \eta_1 +  g \eta_1 = B_1  && \text{at $y=H$}\\
&\psi_1 +  \psi_0'(H) \eta_1= 0 && \text{at $y=H$}\\
&\psi_1=0 && \text{at $y=0$}
\end{aligned}\right.
\]
for some constant $B_1$, where primes denote ordinary differentiation. Seeking solutions of the form
\[
\psi_1(x,y) = f_1(y) \cos(kx) \quad\text{and}\quad \eta_1(x) = A_1\cos(kx)
\]
for some wave number $k$, we arrive at
\begin{subequations}\label{eq:f}
\begin{align}
&f_1'' + (\gamma'(\psi_0)- k^2)f_1= 0 \quad \text{for $0<y<H$}, \label{eq:lin111} \\
&\psi_0'(H) f_1'(H) + (\psi_0'(H) \psi_0''(H) + g) A_1 = 0, \label{eq:lin222} \\
&f_1(H) + \psi_0'(H) A_1= 0, \label{eq:lin333} \\
&f_1(0) =0.   \label{eq:lin444}
\end{align} 
\end{subequations}

To determine the dispersion relation, we proceed as follows. Given $Q$, we solve \eqref{eq:psi0} for $\psi_0(y)$ using a numerical shooting method. Substituting $\psi_0(y)$ into \eqref{eq:f}, we solve \eqref{eq:lin111} for $f_1(y)$ subject to the boundary conditions \eqref{eq:lin333} and \eqref{eq:lin444} using a shooting method. It remains to determine the value of $Q$ such that \eqref{eq:lin222} is satisfied. This can be done graphically by plotting the residual of \eqref{eq:lin222} as a function of $Q$ or alternatively by employing an iterative solver such as the Newton--Raphson method. 

In some special cases, the dispersion relation can be derived analytically. For instance, in an irrotational flow where $\gamma(\psi)=0$, 
\[
\psi_0(y) = \frac{Q}{H}y \quad \text{and} \quad 
f_1(y)= -A_1\frac{Q}{H}\frac{\sinh(ky)}{\sinh(kH)},
\]
which leads to the well-known dispersion relation $c^2 = \frac{g}{k}\tanh(kH)$. 
For constant vorticity, $\gamma(\psi)=\gamma_0$, a straightforward calculation reveals
\[
\psi_0(y) = cy-\frac{\gamma_0}{2}y^2 \quad \text{and} \quad
f_1(y)= -A_1\Big(\frac{c}{H}-\gamma_0\Big)\frac{\sinh(ky)}{H\sinh(kH)},
\]
where $c=\frac{Q}{H} + \frac{\gamma_0}{2} H$. The dispersion relation then follows:
\[
(c-\gamma_0 H )^2 + \frac{\gamma_0}{k}\tanh(kH)(c-\gamma_0 H ) - \frac{g}{k}\tanh(kH)  = 0,
\]
which is consistent with the result, for instance, in \cite{da1988steep}. 

Another instance where an analytical solution is possible is an affine vorticity function, that is,  
\[
\gamma(\psi) = a\psi+b \quad \text{for some $a>0$}\quad \text{for some $b\in \mathbb{R}$}.
\]
We follow the approach of \cite{ehrnstrom2011steady} but with a different scaling. Solving \eqref{eq:psi0}, we obtain
\[
\psi_0(y) = C\cos(\sqrt{a}y) + S\sin(\sqrt{a}y) - \frac{b}{a},
\]
where 
\begin{align}\label{eq:beta}
C&= \frac{b}{a} \quad\text{and}\quad 
S= \frac{Q+\frac{b}{a}(1-\cos(\sqrt{a}H))}{\sin{\sqrt{a}H}}, 
\end{align}
provided $a\neq (n\pi/H)^2$ for $n\in\mathbb{N}$. If $a=(n\pi/H)^2$, then $Q=0$ must hold. For notational convenience, we define
\[
C_H= \cos(\sqrt{a}H) \quad \text{and} \quad S_H= \sin(\sqrt{a}H).
\]
Solving \eqref{eq:lin111} and \eqref{eq:lin444}, we find
\[
f_1(y) = \begin{cases}
F \sin(\delta y), & a>k^2 \\ 
F y, & a=k^2 \\
F \sinh(\delta y),  &a<k^2
\end{cases}
\]
for some constant $F$, where $\delta^2=a-k^2$. To simplify the algebra, we focus on the case $a>k^2$, noting that similar expressions hold for $a\leq k^2$. It follows from \eqref{eq:lin333} that
\[
F = \frac{\sqrt{a}(CS_H-SC_H) }{\sin (\delta H)} A_1.
\]
Noting $f_1'(H) = \delta F \cos(\delta H)$, \eqref{eq:lin222} can be written as a quadratic equation for $S$:
\begin{equation}\label{eq:ldr_affine2}
\begin{aligned}
(\delta C_H \cot(\delta H) + \sqrt{a} S_H ) C_H S^2 -& (2\delta S_HC_H \cot(\delta H) + \sqrt{a} (S_H^2-C_H^2) )CS \\ 
&+ \delta \cot(\delta H) S_H^2C^2-  \sqrt{a} C_HS_H C^2- \frac{g}{a} =0. 
\end{aligned}
\end{equation}
We denote the two roots of this quadratic equation by $S^\pm$. Recalling the latter equation of \eqref{eq:beta}, we obtain the dispersion relation
\begin{equation}\label{eq:ldr_affine}
Q^\pm = S^\pm S_H - C(1-C_H).
\end{equation}
The algebra simplifies considerably when $b=0$, and
\[
Q^2 = \frac{gS_H^2}{aC_H(\delta C_H \cot(\delta H) +  \sqrt{a}S_H )}.
\]
In the long-wave limit as $k\rightarrow 0$, 
\begin{equation}\label{eq:beta_longwave}
\sqrt{a} \cot(\sqrt{a}H) S^2 - \sqrt{a} CS - \frac{g}{a} =0 .
\end{equation}
The dispersion relation forms asymptotes when the coefficient of $S^2$ in \eqref{eq:ldr_affine2} vanishes, that is, when
\begin{equation}\label{eq:asymptote}
\sqrt{a-k^2}\cot(\sqrt{a-k^2}) = -\sqrt{a} \tan(\sqrt{a}H).
\end{equation}
The number of solutions for $k$ to this transcendental equation is zero for sufficiently small $a$, but it increases as $a$ increases. The existence of long waves depends on whether \eqref{eq:beta_longwave} has two real roots, which occurs if
\[
b^2 + 4 \sqrt{a} g \cot(\sqrt{a}H)>0.
\]
See Section~\ref{sec:results} for more details of the dispersion relation. 

\section{Numerical method}\label{sec:method}

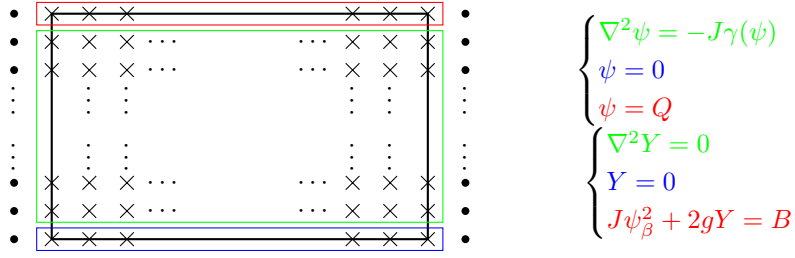
\begin{figure}	
\centering
	\begin{tikzpicture}[domain=-2*pi:2*pi]
		\tikzmath{\Lb=5; \Hb=3; \M=10; \N=8; \Mp=\M+1; \Mm=\M-1; \Mmm=\M-2; \Nm=\N-1; \Nmm=\N-2; \Dx=\Lb/\M; \Dy=\Hb/\N; \bx=.4*\Lb/\M; \by=.4*\Hb/\N;}
		\node (A) at (0,0) {};
		\node (B) at (\Lb,0) {};
		\node (C) at (0,\Hb) {};
		\node (D) at (\Lb,\Hb) {};	
		\draw[draw=black, thick] (0,0) rectangle ++(\Lb,\Hb);
		
		\foreach \i in {0,1,2,\Mmm,\Mm,\M}
		{
			\foreach \j in {0,1,2,\Nmm,\Nm,\N}
			{
				\draw (\Lb*\i/\M,\Hb*\j/\N) node[cross,rotate=0] {};
		}}	

        \foreach \j in {0,1,2,\Nmm,\Nm,\N}
			{
              \node[circle,fill=black,inner sep=0pt,minimum size=3pt] at (\Lb+\Lb*1/\M,\Hb*\j/\N)  {};
               \node[circle,fill=black,inner sep=0pt,minimum size=3pt] at (-\Lb*1/\M,\Hb*\j/\N)  {};
                
            }
            
		\node at (\Lb*2/\M+\Dx,\Hb*1/\N) {$\cdots$};
		\node at (\Lb*2/\M+\Dx,\Hb*2/\N) {$\cdots$};
		\node at (\Lb*2/\M+\Dx,\Hb*\Nmm/\N) {$\cdots$};
		\node at (\Lb*2/\M+\Dx,\Hb*\Nm/\N) {$\cdots$};
		\node at (\Lb*\Mmm/\M-\Dx,\Hb*1/\N) {$\cdots$};
		\node at (\Lb*\Mmm/\M-\Dx,\Hb*2/\N) {$\cdots$};
		\node at (\Lb*\Mmm/\M-\Dx,\Hb*\Nmm/\N) {$\cdots$};
		\node at (\Lb*\Mmm/\M-\Dx,\Hb*\Nm/\N) {$\cdots$};

        \node[rotate=90] at (\Lb*-1/\M,\Hb*2/\N+\Dy) {$\cdots$};
		\node[rotate=90] at (\Lb*1/\M,\Hb*2/\N+\Dy) {$\cdots$};
		\node[rotate=90] at (\Lb*2/\M,\Hb*2/\N+\Dy) {$\cdots$};
		\node[rotate=90] at (\Lb*\Mmm/\M,\Hb*2/\N+\Dy) {$\cdots$};
		\node[rotate=90] at (\Lb*\Mm/\M,\Hb*2/\N+\Dy) {$\cdots$};		
         \node[rotate=90] at (\Lb*\Mp/\M,\Hb*2/\N+\Dy) {$\cdots$};	
        \node[rotate=90] at (\Lb*-1/\M,\Hb*\Nmm/\N-\Dy) {$\cdots$};
        \node[rotate=90] at (\Lb*1/\M,\Hb*\Nmm/\N-\Dy) {$\cdots$};
		\node[rotate=90] at (\Lb*2/\M,\Hb*\Nmm/\N-\Dy) {$\cdots$};
		\node[rotate=90] at (\Lb*\Mmm/\M,\Hb*\Nmm/\N-\Dy) {$\cdots$};
		\node[rotate=90] at (\Lb*\Mm/\M,\Hb*\Nmm/\N-\Dy) {$\cdots$};
        \node[rotate=90] at (\Lb*\Mp/\M,\Hb*\Nmm/\N-\Dy) {$\cdots$};
		\draw[draw=blue] (-\bx,-\by) rectangle ++(\Lb+2*\bx,2*\by);
		\draw[draw=red] (-\bx,\Hb-\by) rectangle ++(\Lb+2*\bx,2*\by);
		\draw[draw=green] (-\bx,\Dy-\by) rectangle ++(\Lb+2*\bx,\Hb-3*\by);
		
		\node at (\Lb + 7*\Dx, 3*\Hb/4){$\begin{cases} {\color{green} \nabla^2 \psi =-J \gamma(\psi)} \\ 
				{\color{blue} \psi = 0} \\
				{\color{red}  \psi = Q}
			\end{cases}$};
		\node at (\Lb + 7.3*\Dx, \Hb/4){$\begin{cases} {\color{green} \nabla^2 Y = 0} \\ 
				{\color{blue} Y = 0} \\
				{\color{red}J \psi_\beta^2 + 2gY = B}
			\end{cases}$};
		
\end{tikzpicture}	
\caption{\label{fig:discr} Schematic of the discretisation of $\Omega_{(\alpha,\beta)}^+$. The field equations and the boundary conditions are applied at mesh points corresponding to the coloured boxes. Crosses are mesh-points, while dots show ghost points, which combined with \eqref{eq:neumann1} and \eqref{eq:neumann2} give second order approximations \eqref{eq:secondorderedge}. }
\end{figure}

We describe how we numerically solve \eqref{eq:psi}-\eqref{eq:Y} using a finite difference scheme. Our approach is similar to previous studies that reformulates the problem through the semi-hodograph transformation \citep{dalrymple1977numerical,thomas1990wave,ko2008large,ko2008effect,amann2018numerical}.

Since solutions are assumed to be symmetric about $\alpha=0$, we restrict computations to the half-domain $\Omega_{(\alpha,\beta)}^+$ where $\alpha\geq 0$. The boundary condition at $\alpha=0$ is given by $Y_\alpha=0$, the same as the condition at $\alpha=\pm\pi$, so that our numerical method is applicable to symmetric and non-symmetric solutions although for non-symmetric solutions, computations span a full wavelength in $\Omega_{(\alpha,\beta)}^+$ rather than a half-wavelength. 

While the method is formulated for periodic waves in finite depth, solitary waves can be approximated by taking the wavelength $L$ sufficiently large such that further increases in $L$ do not result in significant changes in the solution. Similarly, waves in infinite depth can be approximated by selecting a sufficiently large depth $H$.

We discretise $\Omega_{(\alpha,\beta)}^+$ into $M$ equally spaced points in $\alpha$ and $N$ equally spaced points in $\beta$ as
\[
\alpha_m =\pi\frac{m-1}{M-1}\quad\text{and}\quad 
\beta_n =-d \frac{N-n}{N-1},\quad m=1,\cdots,M, \quad n=1,\cdots,N.
\]
The values of $M$ and $N$ are chosen depending on the computed solutions. For instance, for solitary waves approximated by long periodic waves, larger values of $M$ are preferable to accurately resolve horizontal variations, while periodic waves in infinite depth necessitate larger values of $N$ to adequately capture the flow motion at greater depths. Throughout Section~\ref{sec:results}, the values of $M$ and $N$ used in computations are provided. We define 
\[
\Delta\alpha=\frac{\pi}{M-1} \quad \text{and} \quad \Delta\beta = \frac{d}{N-1}.
\]

Let $Y_{m,n}$ and $\psi_{m,n}$ denote the values of $Y$ and $\psi$ at the mesh point $(\alpha_m, \beta_n)$ respectively. Ghost points are introduced at the left and right boundaries, where $\alpha_0=-\Delta\alpha$ and $\alpha_{M+1}=\pi+\Delta\alpha$. Applying the boundary conditions \eqref{eq:neumann1} and \eqref{eq:neumann2} at $(\alpha_1,\beta_n)$ and $(\alpha_M,\beta_n)$, and employing second-order central differences, we obtain $Y_{0,j}=Y_{2,j}$ and $Y_{M+1,j}=Y_{M-1,j}$, and similarly for $\psi$. The second-order central difference approximations at the left and right boundaries lead to
\begin{equation} \label{eq:secondorderedge}
Y_{\alpha\alpha}(\alpha_1,\beta_n) \approx \frac{2(Y_{2,n}-Y_{1,n})}{\Delta \alpha^2}
\quad\text{and}\quad 
Y_{\alpha\alpha}(\alpha_M,\beta_n) \approx \frac{2(Y_{M-1,n}-Y_{M,n})}{\Delta \alpha^2}.
\end{equation}

The field equations are solved at internal mesh points $(\alpha_m,\beta_n)$, where $m=1,\cdots,M$ and $n=2,\cdots,N-1$, using second-order central differences to approximate the derivatives. The boundary conditions \eqref{eq:dirichlet2} and \eqref{eq:dirichlet3} are enforced at $(\alpha_m,\beta_1)$ for $m=1,\cdots,M$. The boundary conditions \eqref{eq:dirichlet1} and \eqref{eq:dbc} are imposed at $(\alpha_m,\beta_N)$ for $m=1,\cdots,M$, where $\psi_\beta$ and $Y_\beta$ are approximated using the second-order backward Euler equation, and $Y_\alpha$, for $m=2,\cdots,M-1$, are approximated using second-order central differences, and set to zero at $m=1$ and $M$. See Figure~\ref{fig:discr}.

This involves $2MN+1$ unknowns---the values of $\psi$ and $Y$ at each mesh point along with the Bernoulli constant $B$---while providing only $2MN$ equations. Additional constraints must be imposed to ensure a unique solution. This can be achieved by fixing bifurcation parameters. One approach is to fix the wavelength 
\begin{align}\label{eq:wavelength}
L = 2\int_{0}^{\pi} Y_\beta(\alpha,\beta)~d\alpha \Big|_{\beta=\text{const.}},
\end{align}
which introduces an additional equation. Alternatively, one can fix the fluid depth
\begin{equation}\label{eq:height}
H=\int_{0}^{\pi} Y(\alpha,0)~d\alpha \quad \text{or} \quad H_0=Y(\pi).
\end{equation}
We introduce $H_0$ because it is computationally less expensive to impose than the nonlocal equation for $H$. Furthermore, for solitary waves, $H_0$ exactly corresponds to the fluid depth.
Fixing the dimension of the mapping space, we can fix only one of $L$ and $H_0$, and treat $Q$ as an unknown. Additionally, we choose a bifurcation parameter, such as the wave amplitude 
\begin{equation}\label{eq:amp}
A = Y_{1,N} - Y_{M,N}.	
\end{equation}
This results in $2MN+2$ unknowns---$\psi_{m,n}$, $Y_{m,n}$, $B$, $Q$---which are matched by $2MN+2$ equations---$2MN$ from the field equations and boundary conditions, \eqref{eq:amp}, and one of \eqref{eq:wavelength} or \eqref{eq:height}. 
If both $L$ and $H$ (or $H_0$) are prescribed, the dimension of the mapping space must be allowed to vary, that is, $d$ is an unknown, leading to  $2MN+3$ unknowns---$\psi_{m,n}$, $Y_{m,n}$, $B$, $Q$, $d$. If neither $H$ nor $L$ is fixed, then both $Q$ and $d$ are prescribed, and there are $2MN+1$ unknowns. Throughout Section~\ref{sec:results}, we specify which physical parameters are fixed for each solution. 

The Newton--Raphson method is employed for numerical solution. Convergence is achieved when the $L^\infty$-norm of the residuals falls below the tolerance
\begin{equation}\label{eq:tol}
\text{TOL} = \frac{10^{-13}}{\max(\Delta \alpha^2,\Delta \beta^2)},
\end{equation}
which accounts for rounding errors introduced by double-precision floating-point arithmetic. Convergence is typically reached within a few iterations. The biggest computational bottleneck is the construction of the Jacobian matrix at each iteration. While the linear part of the Jacobian remains fixed for a given $d$, the nonlinear terms must be updated. Fortunately, the sparsity of the Jacobian matrix allows for efficient inversion. Computational speed can be further improved by performing multiple iterations with the same Jacobian once the residuals are sufficiently small.

\section{Results}\label{sec:results}

Given the wide range of possible vorticity functions, a comprehensive treatment of all solutions is infeasible. Instead, we focus on a selected set of vorticity functions, chosen based on their physical relevance and connections to theoretical studies of rotational flows. 

\subsection{Zero vorticity}

We begin by validating our numerical method against known results for solitary waves in an irrotational flow. To this end, we compare solutions from our finite difference scheme with those obtained using a high-order series truncation method, for instance, in \citep{doak2017solitary}. Taking $g=1$ and $H_0=1$, the far-field fluid velocity $=\lim_{x\rightarrow \infty}|\psi_y(x,y)|$, which independent of $y$, is given by the nondimensional Froude number, denoted $F$. We choose the conformal depth $d$ sufficiently small such that further decreases in $d$ do not alter the computed solutions significantly.

It is well established that small-amplitude solitary waves bifurcate from $F=1$ and the amplitude increases with $F$ until it reaches a limiting configuration which is distinguished by a stagnation point at the crest possessing a $120^{\circ}$ angle. This limiting scenario was conjectured by \citet{stokes1847theory,stokes1880theory} and later proved in \citep{AFT1982} \citep[see also][]{toland1978existence}.

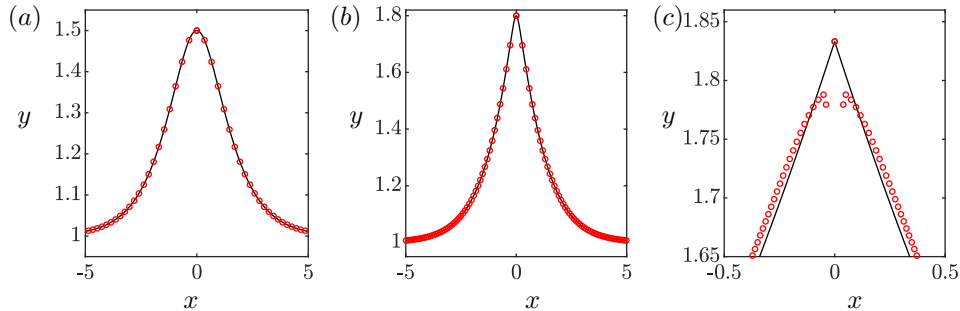
\begin{figure}
\centering
\begin{overpic}[scale=1]{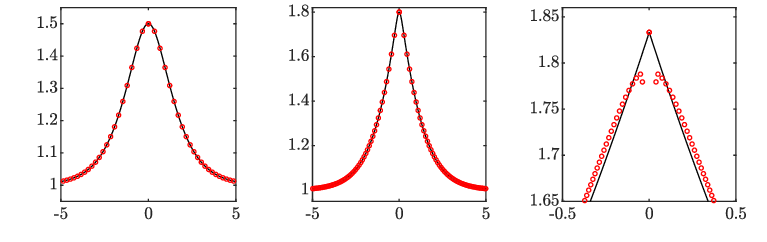}
    \put(0,29){$(a)$}
    \put(33,29){$(b)$}
    \put(66,29){$(c)$}
    \put(18,0){$x$}
    \put(53,0){$x$}
    \put(86,0){$x$}
    \put(1,19){$y$}
    \put(35,19){$y$}
    \put(67,19){$y$}
\end{overpic}
\caption{Solitary wave profiles for zero vorticity computed using the finite difference scheme (red circles) and a high-order method (black curves). Panels $(a)$ and $(b)$ depict solutions for $A=0.5$ and $A=0.8$, showing excellent agreement between the two methods. Panel $(c)$ illustrates a wave of almost greatest height, where the finite difference scheme fails to accurately capture the singularity behaviour with significant unphysical oscillations.}
    \label{fig:irrot1}
\end{figure}

Figure~\ref{fig:irrot1} presents three wave profiles. Panels $(a)$ and $(b)$ show solutions with amplitudes $A=0.5$ and $A=0.8$, respectively, demonstrating excellent agreement between the result from our finite difference scheme (red circles) and that from the series truncation method (solid black curves). However, panel $(c)$ shows the poor performance of the finite difference scheme near the wave of greatest height, where $A\approx 0.8332$. It does not adequately resolve the singular behaviour near the crest, and spurious oscillations appear in the approximations of the derivatives. 

\begin{figure}
\centering
\begin{overpic}[scale=1]{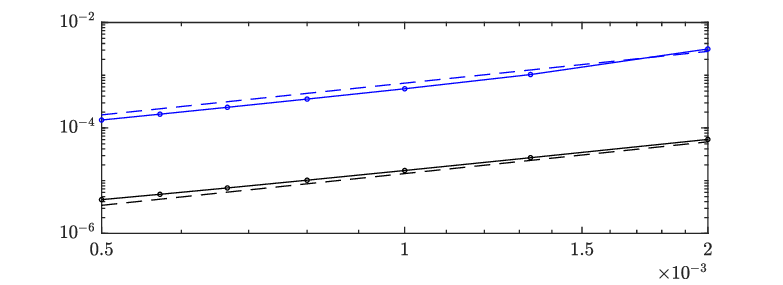}
    \put(3,15){\rotatebox{90}{$|F-F_\text{true}|$}}
    \put(50,0){1/M}
\end{overpic}
\caption{Log-log plot of the error in the computed Froude number as a function of mesh sizing. For all solutions, $N=M/5$. The value $F_\text{true}$ is obtained a series truncation method. The black and blue curves correspond to the solutions shown in Figure~\ref{fig:irrot1} $(a)$ and $(b)$. The dashed lines indicate quadratic convergence.}
\label{fig:ERR}
\end{figure}

Figure \ref{fig:ERR} presents a log-log plot of the error in the computed Froude number $F$ versus the number of mesh points for $A=0.5$ (solid black curve) and $A=0.8$ (solid blue curve). The `true' value of $F$ is taken from the result using the series truncation method. Setting $M=5N$, we find that as $M$ and, hence, $N$ increase, the convergence of the numerical result is approximately quadratic in $1/M$. This is expected because of the use of second-order differences throughout the discretisation. 

\subsection{Constant vorticity}\label{section:constant}

To further assess the accuracy and capability of our numerical scheme, we consider constant vorticity, for which \eqref{eq:psi} is no longer trivially satisfied. 

\begin{figure}
    \centering
    \begin{overpic}[scale=1]{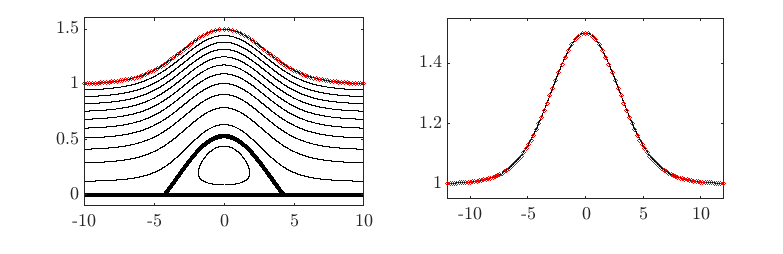}
    \put(28,1){$x$}
    \put(3,17){$y$}
    \put(74,1){$x$}
    \put(50,17){$y$}
    \put(3,28){$(a)$}
    \put(50,28){$(b)$}
    \end{overpic}
    \caption{Solitary wave for constant vorticity $\gamma=5$ for $g=1$, $H_0=1$, and $A=0.5$. Panel $(a)$ presents streamlines, while panel $(b)$ displays only the fluid surface. The black curves represent the result from our numerical method, while the red circles are computed in \citep{guan2020particle}. The mesh sizes are $M=2000$ and $N=400$. The bold black curves denote the fluid bed and a streamline that forms a critical layer.}\label{fig:guan}
\end{figure}
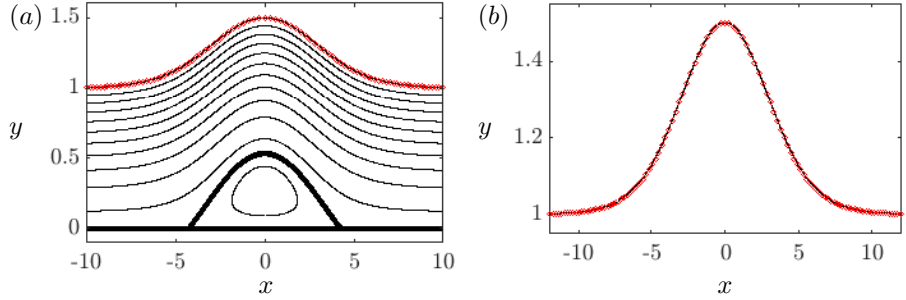

We begin by examining a solitary wave for $\gamma=5$. We take $g=1$, $H_0=1$ and $A=0.5$. Figure~\ref{fig:guan} presents the wave profile computed using our finite difference scheme (black curves) alongside the result from a different numerical approach (red circles) solving for the conformal mapping of the irrotational correction to a linear shear current, for instance, in \cite{guan2020particle}. Panel $(b)$ displays the fluid surface, confirming strong agreement between the two methods. Panel $(a)$ illustrates the streamlines throughout the fluid, revealing a critical layer with two stagnation points along the fluid bottom (bold black curves). Notably, such a solution would not be possible to compute using the semi-hodograph transformation.

\begin{figure}
    \centering
    \begin{overpic}[scale=1]{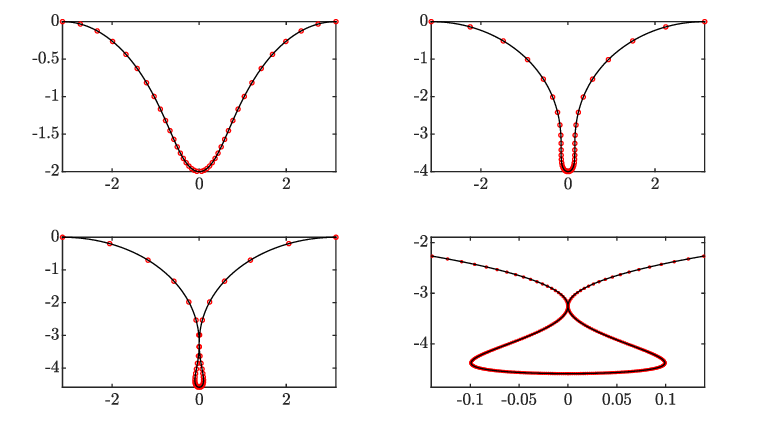}
    \put(0,52){$(a)$}
    \put(48,52){$(b)$}
    \put(0,24){$(c)$}
    \put(48,24){$(d)$}
    \put(25,27){$x$}
    \put(72,27){$x$}
    \put(25,0){$x$}
    \put(72,0){$x$}
    \put(1,43){$y$}
    \put(49,43){$y$}
    \put(1,15){$y$}
    \put(49,15){$y$}
    \end{overpic}
    \caption{Periodic travelling waves in infinite depth for $\gamma=1$ and $g=0$. The black curves represent Crapper's exact solutions, taken from \citep{hur2020exact}. The red circles denote solutions computed using our numerical scheme, taking $L=2\pi$ and $d=7.5$. Not all mesh points are displayed. Panels $(a)$ and $(b)$ correspond to $A=2$ and $A=4$ (see \eqref{eq:crapper}). Panels $(c)$ and $(d)$ illustrate the wave of maximum amplitude, with panel $(d)$ providing a close-up of a touching region. The $y$-axis has been shifted such that the interface is at $y=0$ when $x=\pm\pi$. For all numerical solutions, $M=200$ and $N=3000$. }
    \label{fig:crapper}
\end{figure}
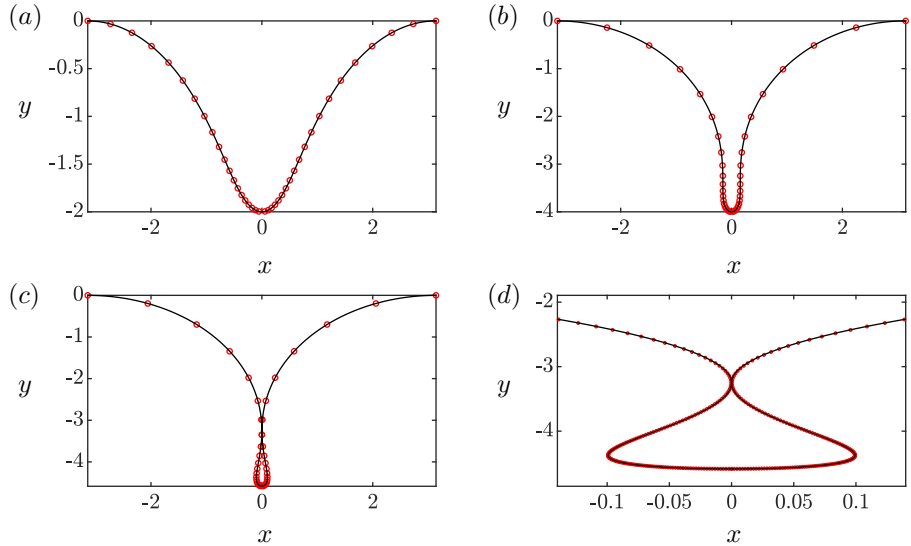

Also, we consider zero gravity and infinite depth. \citet{crapper1957exact} derived an exact solution for capillary waves in an irrotational flow of infinite depth, and a remarkable recent result established the equivalence between Crapper's wave profiles and those in constant vorticity flows in infinite depth in the absence of surface or body forces. This was hypothesised based on numerical and asymptotic evidence \citep{dyachenko2019stokes,hur2020new} and rigorously proved in \cite{hur2020exact}. 
For $\gamma=1$ and $L=2\pi$ ($d=\infty$ and $g=0$), the free surface can be parameterised as $(X(\alpha),Y(\alpha))$, $-\pi\leq \alpha\leq \pi$, where
\begin{align}\label{eq:crapper}
X(\alpha) + i Y(\alpha) &= \alpha - \frac{4iC}{C+e^{i\alpha}},
\end{align}
where $C\in[0,C_\textit{max}]$ serves as an amplitude parameter. When $C=0$, the wave has zero amplitude ($A=0$).  When $C=C_{\max}\approx0.4547$, on the other hand, the wave attains its maximum amplitude ($A_{\max}\approx 4.5851)$, for which the fluid surface intersects itself tangentially above the trough, enclosing an air bubble. 

In Figure~\ref{fig:crapper}, we compare solutions computed using our finite difference scheme with the exact solutions in \eqref{eq:crapper}. While we cannot take true infinite depth, we approximate this by fixing the wavelength $=2\pi$ and choosing a sufficiently large $d$ such that further increases in $d$ do not affect the numerical solution. 
For Figure~\ref{fig:crapper}, $d=7.5$. We take $M=200$ and $N=3000$. The solid black curves represent the exact solutions, while red dots denote our numerical result. Note that not all mesh points are displayed. Panels $(a)$ and $(b)$ depict solutions for amplitudes $A=2$ and $A=4$, respectively. Panels $(c)$ and $(d)$ illustrate the wave of maximum amplitude, with the latter providing a close-up of a touching region. In all cases, the numerical and analytical solutions exhibit excellent agreement, further validating the accuracy of our numerical scheme.


\subsection{Affine vorticity function}

We turn the attention to non-constant vorticity functions, with a particular emphasis on solutions that have thus far been numerically unattainable. The parameter space here is vast because the vorticity function can be arbitrary. Motivated by prior studies, we focus on two special cases: affine vorticity functions and two-layer constant vorticity.

\begin{figure}
    \centering
    \begin{overpic}[scale=1]{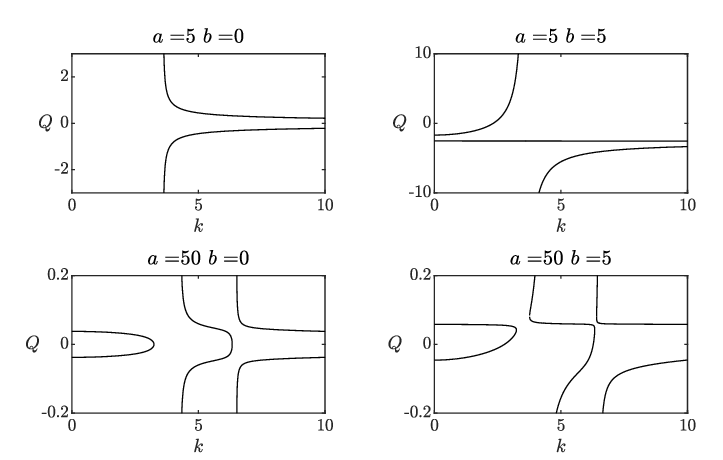}


    \end{overpic}
    \caption{The dispersion relation for $\gamma(\psi)=a\psi + b$ for different values of $a$ and $b$.}
    \label{fig:disprel_affine}
\end{figure}
\begin{figure}
\centering
\begin{overpic}[scale=1]{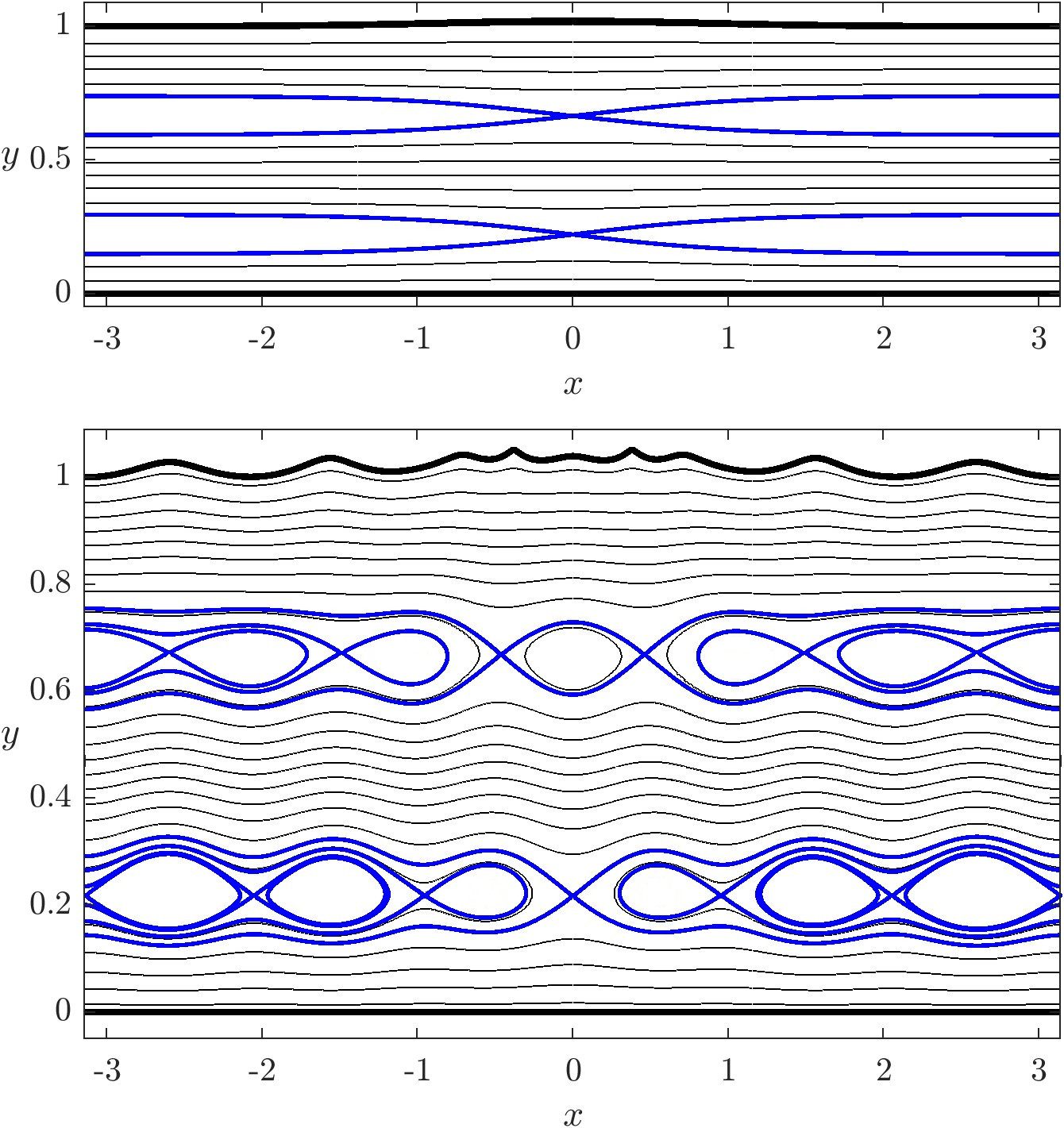}
\put(0,96) {$(a)$}
\put(0,57) {$(b)$}
\end{overpic}
\caption{Solutions for $\gamma(\psi)=50\psi$ and $k=1$, $Q=0.0413$ $(a)$ and $Q=0.0499$ $(b)$. All curves represent streamlines. The bold black lines denote the rigid wall and the fluid surface, while the bold blue lines indicate streamlines with saddle points. The solution in panel $(a)$ is close to linear and exhibits two critical layers with saddle points at $x=0$. The solution in panel $(b)$ approaches a limiting configuration with stagnation points at the fluid surface. There are multiple critical layers inside the flow, revealing an intricate flow structure caused by wave resonances. The solutions have $M=500$ and $N=200$.} \label{fig:a=50_b=0}
\end{figure}
We begin with affine vorticity of the form 
\[
\gamma(\psi)=a\psi+b \quad \text{for some constants $a$ and $b$},
\]
which can admit an arbitrary number of critical layers \citep{ehrnstrom2012steady}. More specifically, examining the solution $\psi_0(y)$ of \eqref{eq:psi0}, we observe that the velocity of the background shear flow oscillates in $y$ with frequency $\sqrt{a}$. When exploring nonlinear solutions, it is helpful to first examine the dispersion relation in the linear theory. Figure~\ref{fig:disprel_affine} presents the dispersion relations for different values of $a$ and $b$. As $a$ increases, the number of asymptotes in $Q(k)$ increases, their locations determined by \eqref{eq:asymptote}. As seen in panels (a) and (c) (see also \eqref{eq:ldr_affine}), $b=0$ results in the two solution branches with opposite values of $Q$. Indeed, when $b=0$, the equations are invariant under $\psi\mapsto -\psi$ and $Q\mapsto -Q$, and the corresponding nonlinear solutions enjoy the same symmetry. 

The asymptotes in the dispersion relation and, hence, multiple values of $k$ for the same value of $Q$, suggests wave resonance. This occurs asymptotically  when $Q(k)=Q(nk)$ for a longer wavelength mode with the wave number $k$ and for a shorter $nk$ mode for some integer $n\geq 2$. Mathematical studies of resonance for capillary-gravity waves in an irrotational flow dates back to \cite{wilton1915lxxii}.

Throughout this subsection, we set the length and time scales such that $g=1$ and $H=1$, unless stated otherwise.

We begin by taking $\gamma(\psi) =50 \psi$ and compute the solution branch for $k=1$. Figure~\ref{fig:a=50_b=0} presents the result. Panel $(a)$ shows a small-amplitude solution, for which there are two critical layers inside the fluid. As the amplitude increases, wave resonances occur, resulting in additional local maxima and minima in the surface displacement. These resonances lead to intricate flow behaviours, including the formation of nested critical layers, as observed in a large-amplitude solution in panel $(b)$. 

Our numerical result suggests that the limiting solution likely exhibits a stagnation point at the wave crest. However, as discussed earlier, our numerical method does not adequately resolve stagnation points at the fluid surface because the conformal mapping, more specifically its second-order derivatives, would become singular there. On the other hand, we observe that the fluid velocity near the crest decreases as one moves along the solution branch, supporting our conjecture that the solution branch would terminate at a solution with a stagnation point at the crest. 

\begin{figure}
\centering
\begin{overpic}[scale=1]{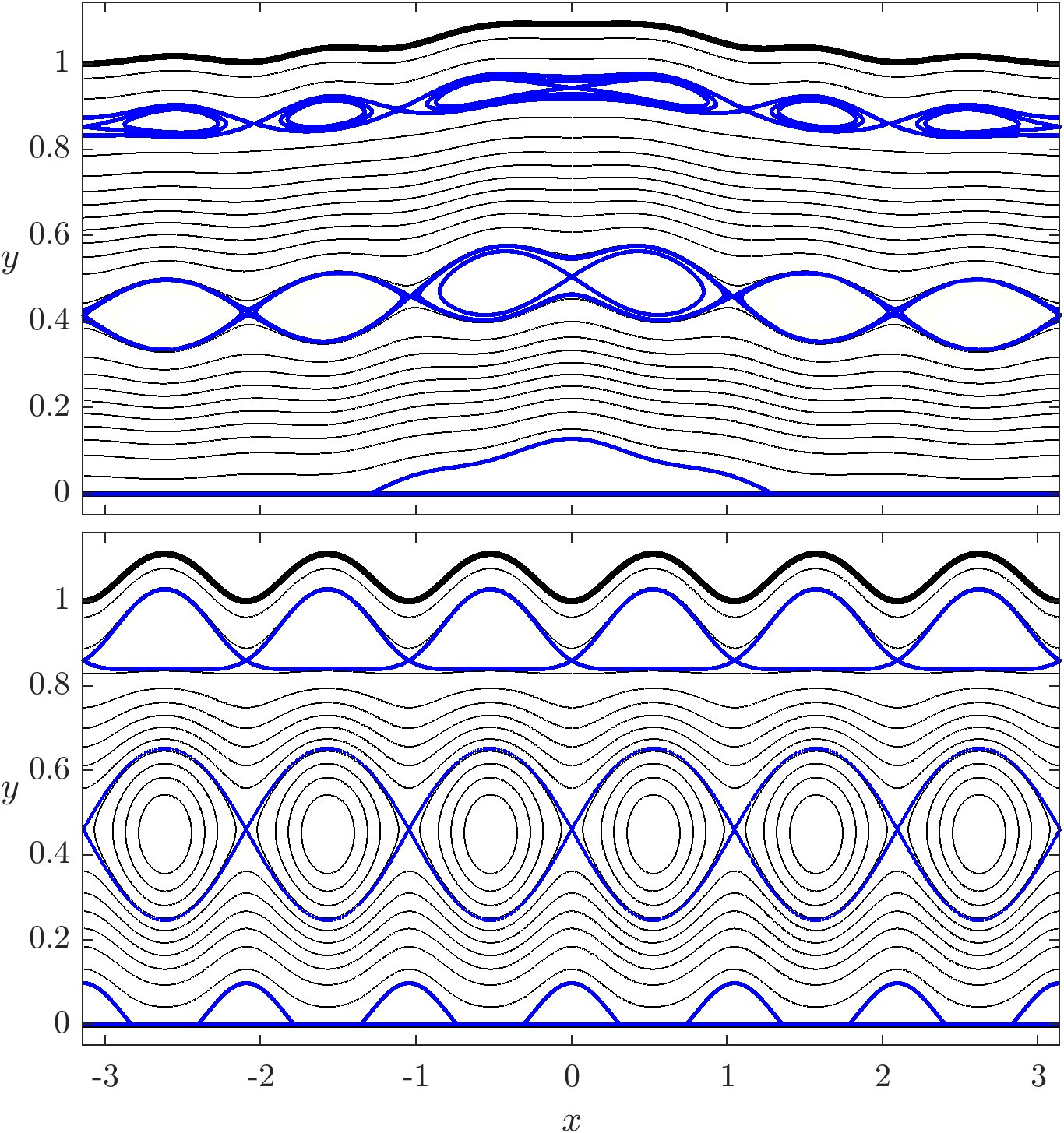}
\put(0,96) {$(a)$}
\put(0,52) {$(b)$}
\end{overpic}
\caption{Solutions for $\gamma(\psi)=50\psi+50$ and $k=1$, $Q=-0.4859$ $(a)$ and $Q=-0.5537$ $(b)$. The solution in panel $(a)$ resembles Wilton ripples. The $k=1$ solution branch connects to a solution for $k=6$, shown in panel $(b)$. The solutions have $M=500$ and $N=200$.}
\label{fig:a=50_b=50}
\end{figure}

This is not the only limiting scenario by which solution branches terminate. For instance, taking $a=50$, $b=50$ and $k=1$, we compute the solution branch bifurcating from $Q=-0.2498$. Figure \ref{fig:a=50_b=50} shows two representative solutions along the branch. In panel $(a)$, wave resonances occur, resulting in multiple local maxima and minima at the fluid surface. As one follows the solution branch, it reaches the solution shown in panel $(b)$, which lies on a solution branch with six periods in the computational domain. In other words, the $k=1$ solution branch reaches a bifurcation point along the $k=6$ solution branch.

\begin{figure}
\centering
\begin{overpic}[scale=1]{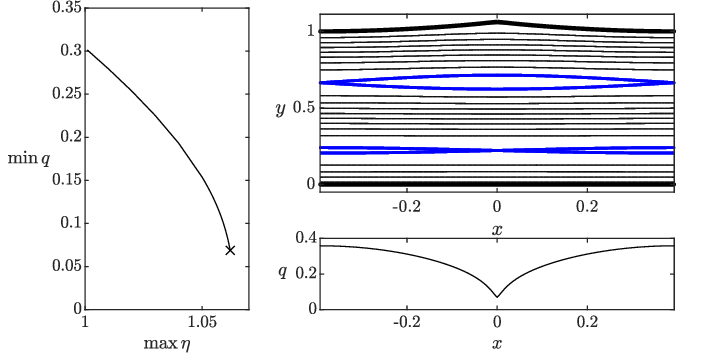}
\put(0,47) {$(a)$}
\put(36,47) {$(b)$}
\put(36,14) {$(c)$}
\end{overpic}
\caption{Solution branch for $\gamma(\psi)=50\psi$ and $k=8$. Panel $(a)$ displays the branch plotted in the $(\min{q}, \max{\eta})$ plane. Bifurcating from an undisturbed interface for $\max{\eta}=1$, the amplitude increases monotonically as the crest speed decreases. Panels $(b)$ and $(c)$ depict the solution marked by the cross in panel $(a)$, representing the farthest point along the branch where solutions could be obtained. Panel $(b)$ shows the streamlines in the physical space, while panel $(c)$ shows the fluid velocity along the free surface as a function of $x$. The solutions were computed with $M=N=500$.} \label{fig:a=50_b=0_k=8}
\end{figure}

For a given $a$ and $b$, wave resonances can be avoided by choosing $k$ sufficiently large. For instance, Figure~\ref{fig:a=50_b=0_k=8} presents solutions with $a=50$, $b=0$ and $k=8$. As shown in Figure \ref{fig:disprel_affine}$(c)$, this choice of $k$ ensures there are no resonant waves number $k>8$ for the same value of $Q$. Under this condition, the wave amplitude increases monotonically along the branch, and the limiting solution exhibits a stagnation point at the crest. The critical layers here exhibit a far less complicated behaviour, forming Kelvin's cat-eye streamline pattern, similar to that observed for constant vorticity \citep{ribeiro2017flow}.

Let $q(x,\eta(x))$ denote the speed of the fluid at point $(x,\eta(x))$. We observe that for all solutions on the non-resonant branch, $\min{q}$ occurs at $x=0$. Figure~\ref{fig:a=50_b=0_k=8}$(a)$ depicts the solution branch in the ($\min{q}$,$\max{\eta}$) space, while panel $(b)$ presents the wave profile in the physical space for the solution at the farthest point along the branch. In panel $(c)$, we plot $q$ as a function of $x$ for the solution in panel $(b)$. As we attempt to further increase the wave amplitude, the near-singular behaviour at the crest results in spurious oscillations, like those observed in an  irrotational flow in Figure~\ref{fig:irrot1}$(c)$.

The amplitude of the near-limiting wave in Figure \ref{fig:a=50_b=0_k=8} is much smaller than that for zero vorticity for the same parameter values for $g$, $H$ and $k$. For constant vorticity, numerical studies \citep{DHS2023} suggest that the limiting wave amplitude for negative vorticity is smaller than that for zero vorticity, whereas the amplitude for positive constant vorticity is larger. Furthermore, for positive constant vorticity, the wave amplitude can grow significantly larger than the irrotational counterpart.
For instance, there are solution branches which approach a limiting configuration consisting of an infinite series of vertically stacked fluid bubbles in rigid body rotation \citep{vanden1996periodic,dyachenko2019stokes}. This raises an interesting question as to what limiting amplitudes are achievable for affine vorticity functions. 

\begin{figure}
\centering
\begin{overpic}[scale=1]{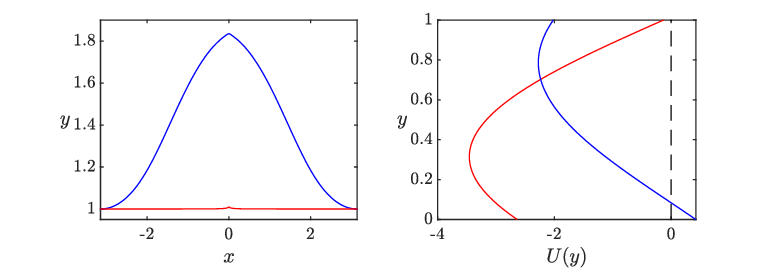}
\put(3,30) {$(a)$}
\put(48,30) {$(b)$}
\end{overpic}
\caption{Solutions for $\gamma(\psi)=5\psi+5$ and $k=1$. Panel $(a)$ displays the wave profiles of near-limiting solutions along the branches bifurcating from $Q=-2.55$ (red) and $Q=-1.47$ (blue). Panel $(b)$ shows the background flow velocity. The red solution attains a near-limiting wave for a very small amplitude, where the background flow velocity is close to zero. The blue solution reaches a much larger amplitude, corresponding to the background flow near its maximum velocity at the fluid surface. The solutions were computed with $M=1000$ and $N=500$.}\label{fig:a=5_b=5.eps}
\end{figure}

We can make some predictions about the limiting amplitude given the background shear flow about which small-amplitude waves bifurcate. To illustrate this, we compute two branches of solutions for $a=5$, $b=5$ and $k=1$, and present our findings in Figure \ref{fig:a=5_b=5.eps}. Nontrivial solutions are found to bifurcate from $Q=-1.47$ and $Q=-2.55$. In panel $(b)$, we show the background flow velocity  $\psi_0'(y)$ for $Q=-1.47$ (blue) and $Q=-2.55$ (red). In panel $(a)$, we present the near-limiting solutions along these branches. We observe that the red solution reaches a limiting solution for a significantly small amplitude, where the background flow velocity is nearly zero at the fluid surface, whereas the limiting amplitude for the blue solution is much larger, corresponding to the background flow velocity near its maximum at the fluid surface. Given the oscillatory nature of the background shear flow, our results suggest that larger wave amplitudes are achieved when the maximum velocity is located close to the fluid surface.    

\begin{figure}
    \centering
    \begin{overpic}[scale=1]{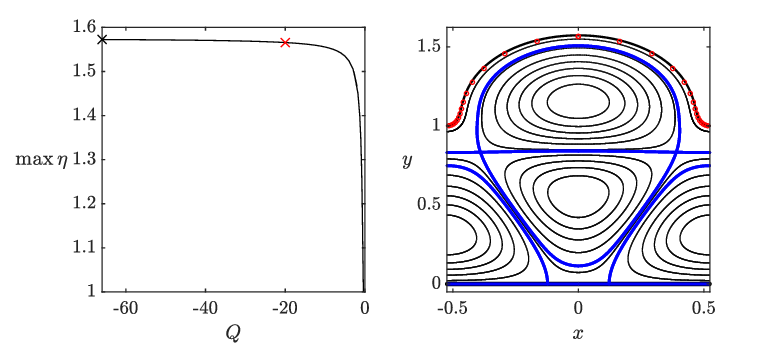}
    \put(3,38){$(a)$}
    \put(49,38){$(b)$}
    \end{overpic}
    \caption{Solutions for $\gamma(\psi)=50\psi+50$ and $k=6$. Panel $(a)$ presents the solution branch in the $(Q,\max{\eta})$ space. Starting from a linear wave with $Q=-0.385$, the wave amplitude increases as $Q$ decreases. Further along the branch, minimal changes in the streamlines occurs as $Q$ decreases further. Panel $(b)$ shows the solution marked by the black cross in panel $(a)$. Streamlines are depicted with the streamlines containing stagnation points. The red dots indicate the free surface of the solution marked by the red cross ($Q=-20$) in panel $(a)$. The solutions were computed with $M=750$ and $N=500$.}
    \label{fig':a=50_b=50}
\end{figure}

Interestingly, this is not the only limiting scenario observed for non-resonant waves. For instance, we take $a=50$, $b=50$, and $k=6$, and examine the branch of solutions bifurcating from $Q=-0.385$. One such solution along this branch, shown in Figure \ref{fig':a=50_b=50}$(b)$, features six wave periods. As the amplitude increases, the branch exhibits an increase in the magnitude of $Q$ with almost no change in the overall flow structure. We plot this branch in the $(Q,\max{\eta})$ space in panel $(a)$. Computations were carried out up to $Q=-65$ with $M=750$ and $N=500$, beyond which further computations became impractical, requiring ever smaller steps in numerical continuation to achieve convergence. Additionally, for large values of $Q$, we had to relax the numerical tolerance by replacing $10^{-13}$ by $10^{-11}$ in \eqref{eq:tol}. We attribute this to an increase in the order of the derivatives of $\alpha$ and $\beta$, with $\nabla^2 \psi$ reaching values of the order of $10^{4}$. In this sense, the relative error remains comparable to that of small-amplitude solutions. In panel $(b)$, we show the largest amplitude solution computed, with $Q=-65$. The red circles show the fluid surface and streamlines with saddle points for the solution with $Q=-25$, illustrating that there is little difference in the fluid surface and flow between these two solutions. 

We have thus far focused on the case of $a>0$. This 

We note that all the limiting behaviour identified through our numerical computations are consistent with the theoretical predictions in \cite{wahlen2023global}.

\subsection{Two-layer constant vorticity}

\begin{figure}
\centering
\begin{overpic}[width=13cm]{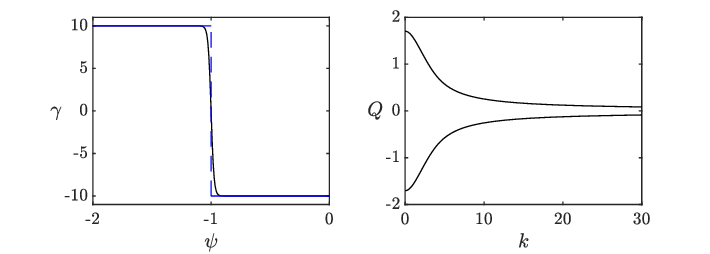}
\put(3,31){$(a)$}
\put(50,31){$(b)$}
\end{overpic}
\caption{$(a)$ The vorticity functions in \eqref{eq:gamma1} (blue) and \eqref{eq:gamma2} (black). $(b)$ The dispersion relation for $g=9.8$, $H=0.6$ for \eqref{eq:gamma2}.}  \label{fig:2layer_dr}
\end{figure}
\begin{figure}
\centering
\begin{overpic}[width=13cm]{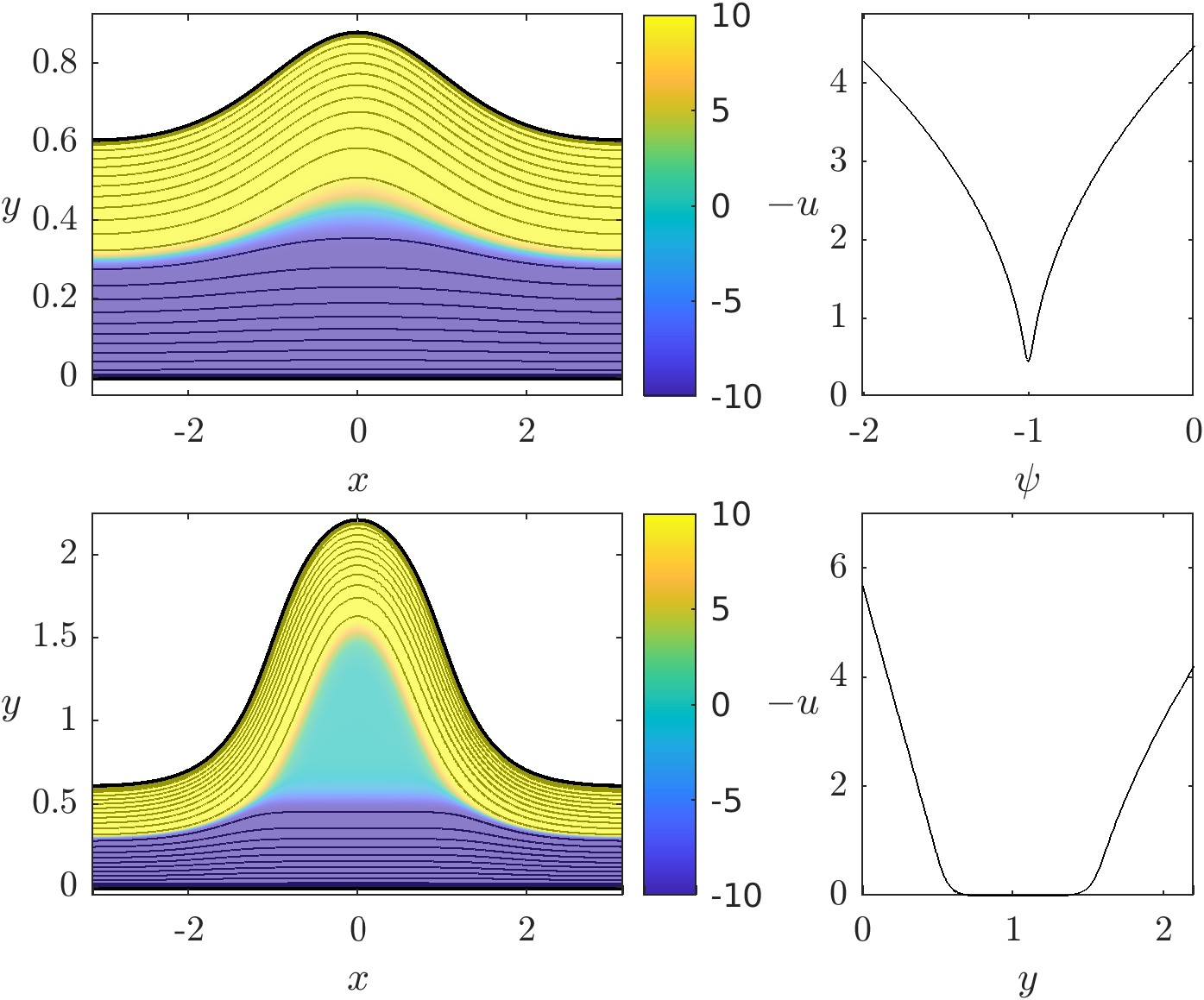}
\put(-2,79){$(a)$}
\put(-2,36){$(c)$}
\put(64,79){$(b)$}
\put(64,36){$(d)$}
\end{overpic}
\caption{Solutions for \eqref{eq:gamma2} for $g=9.8$, $H=0.6$, and $k=1$, $Q=-2$ $(a)$ and $Q=-3.25$ $(c)$. Black lines indicate streamlines, and the colourbar represents vorticity. Panels $(b)$ and $(d)$ show the horizontal velocity at $x=0$ for the solutions in $(a)$ and $(c)$. In $(b)$, the horizontal velocity is plotted as a function of $\psi$ for comparison with \citep{ko2008effect}, and in $(d)$, as a function of $y$ to highlight the broad region of near-stagnation.  The solutions were computed with $M=750$ and $N=500$ \label{fig:2layer}}
\end{figure}
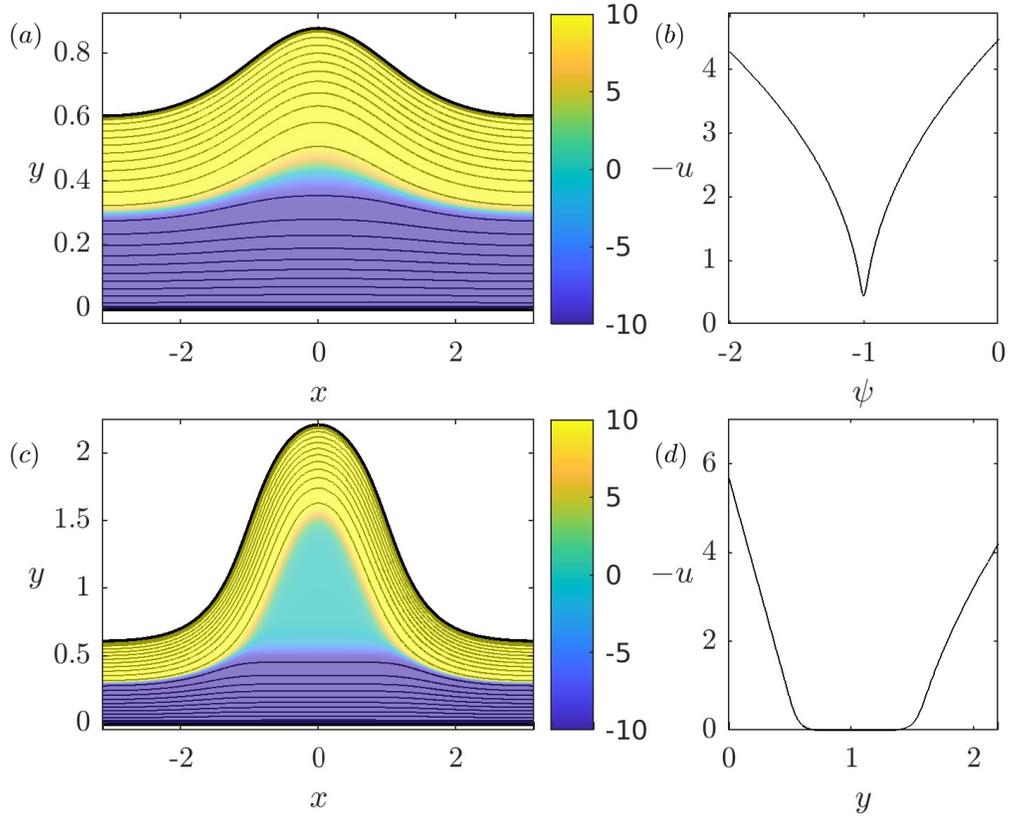

Last but not least, we examine two layers of constant vorticity. Numerical computations for such a vorticity distribution can be found in \cite{ko2008effect, ko2008large}. Figure 12 in \cite{ko2008effect} 
concerns (after adjusting for differences of the choice of where $\psi=0$) solutions for
\begin{align}\label{eq:gamma1}
\gamma(\psi) &= \begin{cases}
10, &-2<\psi<-1 \\
-10, & -1<\psi<0.
\end{cases} 
\end{align}
This creates two layers of constant vorticity flows, with $\gamma=10$ in the upper layer and $\gamma=-10$ in the lower layer, introducing a discontinuity in the vorticity and, hence, in $\nabla^2 \psi$. Since our computations involve second derivatives of $\psi$ using finite differences, we instead seek a smooth vorticity function
\begin{align}\label{eq:gamma2}
 \gamma(\psi) = 10 \tanh\Big(-40\Big(\psi-\frac{Q}{2}\Big)\Big).
\end{align}

Figure \ref{fig:2layer_dr}$(a)$ compares these two vorticity functions for $Q=-2$, illustrating that \eqref{eq:gamma2} provides a smooth approximation of \eqref{eq:gamma1}. To compare with Figure 12 in \cite{ko2008effect}, we fix $g=9.8$, $H=0.6$ and $k=1$. The dispersion relation is shown in Figure \ref{fig:2layer_dr}$(b)$. At $k=1$, we compute solutions bifurcating from $Q=-1.573$. As the amplitude increases, the value of $Q$ decreases, and Figure \ref{fig:2layer}$(a)$ shows the solution obtained at $Q=-2$, in excellent agreement with the result in \citep{ko2008effect}. 

\citet{ko2008effect} employ the semi-hodograph transformation, computing no further along the solution branch once the flow is at near stagnation inside the flow domain, which results in an almost singular change of variables. Figure \ref{fig:2layer}$(b)$ demonstrates this,  where the negative velocity at $x=0$ is plotted as a function of $\psi$, reproducing the result of \citep{ko2008effect}. They conjecture that a stagnation point appears in the interior of the flow at the crest line as the amplitude is further increased. 

Our solution method, which accommodates internal stagnation, allow us to compute solutions beyond interior stagnation. Panel $(c)$ presents a large-amplitude solution, with the colourbar referring to the vorticity distribution. Notably, the region of zero vorticity, occupying only a small region at $x=\pm\pi$, expands to occupy a significant portion of the core of the wave. Panel $(d)$ plots $-u(x=0)$ as a function of $y$, revealing that the velocity nearly vanishes in this region. Continuing further along the branch, the amplitude continues to grow, and the near-stagnant bubble likewise grows in area. 

This raises an intriguing question: what happens in the limiting solution as the vorticity function approaches \eqref{eq:gamma1}?
Does this zero-vorticity region limit to a bubble inside a critical layer? A dedicated study of the two-layer vorticity configuration would be need to explore this possibility.

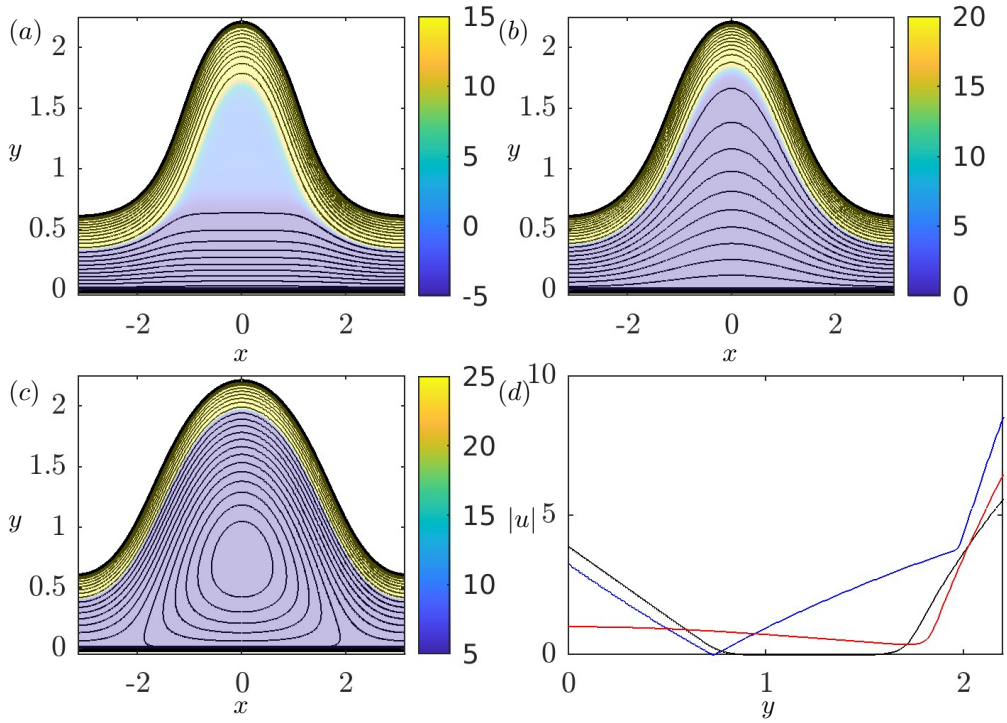
\begin{figure}
\centering
\begin{overpic}[width=13cm]{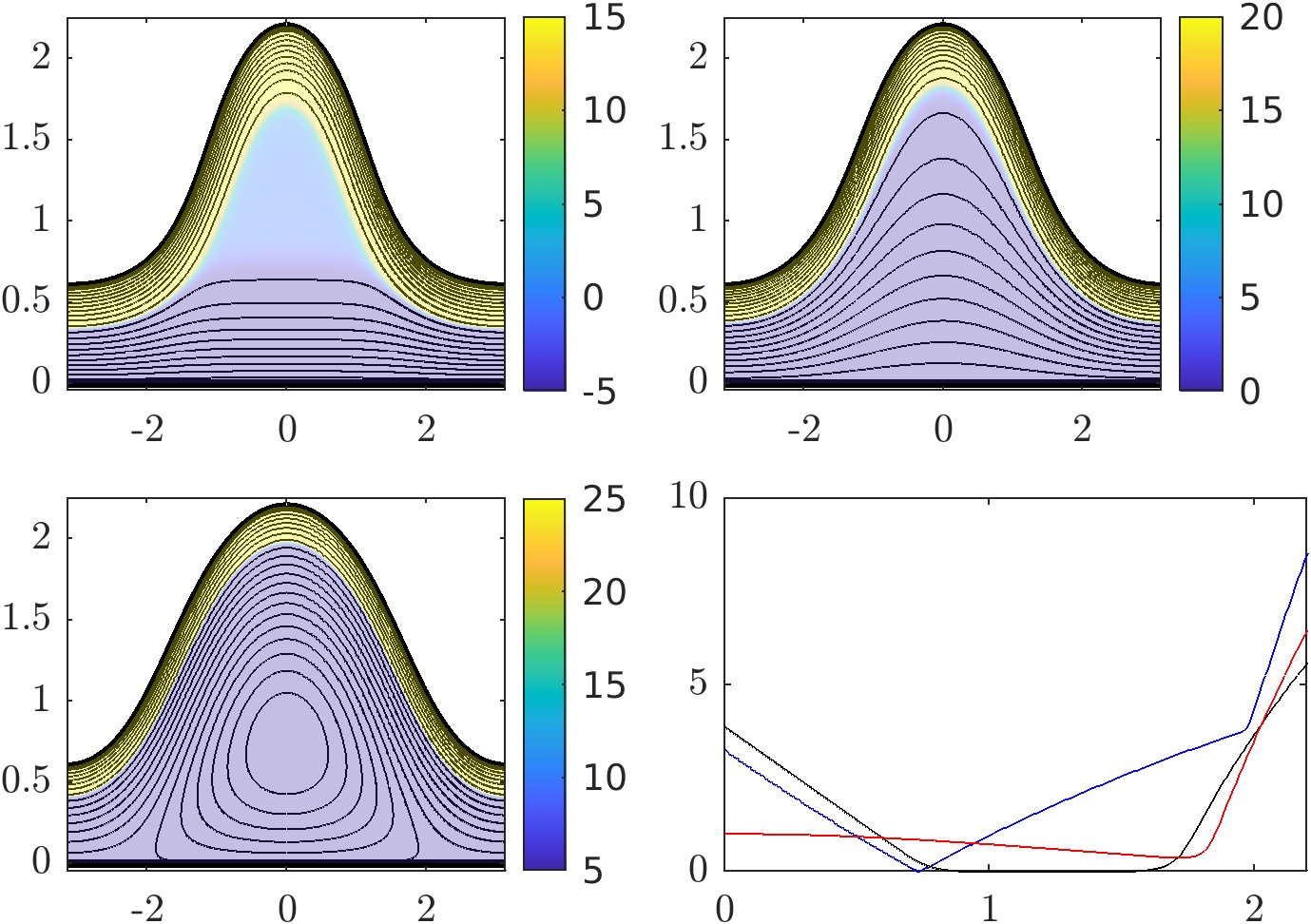}
\put(-2,67){$(a)$}
\put(-2,30){$(c)$}
\put(48,67){$(b)$}
\put(48,30){$(d)$}

\put(-2,55){$y$}
\put(-2,17){$y$}
\put(49,55){$y$}
\put(49,17){$|u|$}

\put(21,-2){$x$}
\put(21,34){$x$}
\put(75,-2){$y$}
\put(71,34){$x$}

\end{overpic}
\caption{Solutions for \eqref{eq:gamma3} for $g=9.8$, $H=0.6$ and $k=1$, where $\gamma_0=5$ $(a)$, $10$ $(b)$, and $15$ $(c)$. Panel $(d)$ shows the magnitude of the horizontal velocity at the crest line, where the black, blue, and red curves correspond to the solutions in $(a)$, $(b)$ and $(c)$.  The solutions were computed with $M=750$ and $N=500$} 
\label{fig:2layer_2}
\end{figure}

The limiting solution of this branch remains unknown. As the amplitude increases further, our numerical method fails to successfully resolve the flow near the crest due to the stretching of mesh points by the conformal mapping. This highlights another limitation of conformal mappings: depending on the wave configuration and, hence, fluid domain, a conformal mapping from a rectangle with the discretisation with equally spaced mesh points can lead to poor resolution in some regions. 

We compared our result with Figure 16 in \cite{ko2008large} but found that even with a large number of mesh points, $M=1300$ and $N=500$, the spacing near the crest became too spread out to maintain accuracy. For some solutions, solving in the $(x,\psi)$ plane may provide better resolution. This issue could potentially be addressed with adaptive meshes or alternative coordinate mapping. We leave this for future work. 

Another interesting observation is that the near-stagnant region appears only when there is a sign change in vorticity. For instance, Figure \ref{fig:2layer_2} presents three solutions for the vorticity function
\begin{equation}\label{eq:gamma3}
\gamma(\psi) = 10 \tanh\Big(-40\Big(\psi-\frac{Q}{2}\Big)\Big) + \gamma_0.
\end{equation}
Panels $(a)$, $(b)$ and $(c)$ correspond to $\gamma_0=5$, $10$ and $15$, respectively. The near-stagnant region is present in panel $(a)$ but absent in panels $(b)$ and $(c)$.

\begin{figure}
    \centering
    \includegraphics[width=11cm]{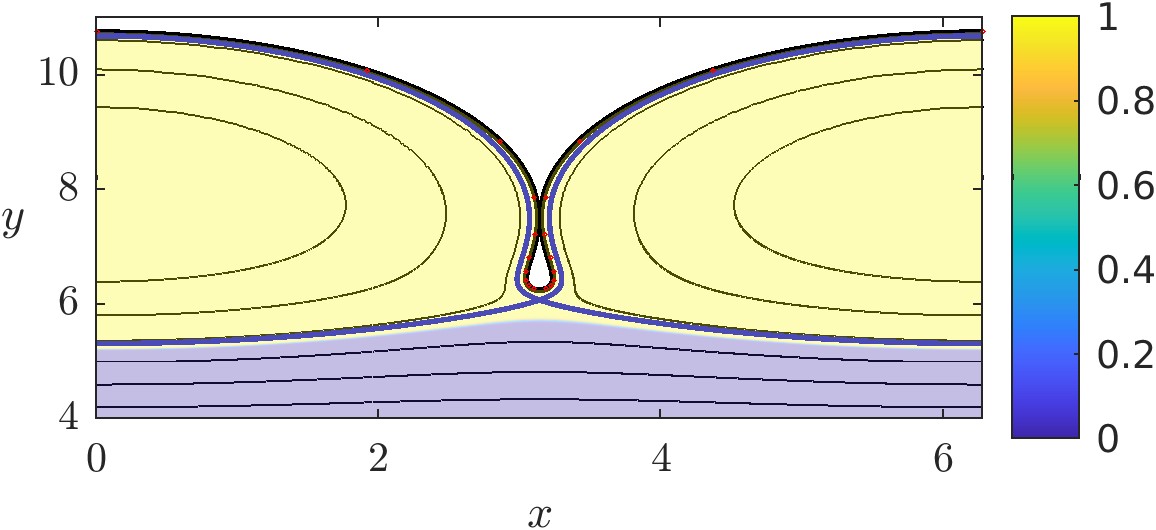}
    \caption{\label{fig:overhang} Solution for \eqref{eq:gamma4} with $g=0$, $d=7.5$, $k=1$ and $Q=8.551$. The lower boundary is given by $y=0$. Black lines indicate streamlines, the blue streamline has a stagnation point that meets at a saddle, and the colourbar represents vorticity. The profile remains largely unchanged from the case of constant vorticity throughout the whole fluid  (shown with red circles) .}
\end{figure}

Finally, we revisit the zero gravity constant vorticity solution computed in section \ref{section:constant}. However, instead of constant vorticity, we consider a two-layer distribution of vorticity, connecting an irrotational bottom layer with a upper layer of constant vorticity of value unity. For this end, we choose
\begin{align}\label{eq:gamma4}
    \gamma(\psi) &= \frac{1}{2} \lrr{1+\tanh\lrr{40\lrr{\psi-\frac{Q}{2}}}}.
\end{align}
Fixing $k=1$ and $d=7.5$, we show a large amplitude solution in figure \ref{fig:overhang}. The free-surface shape remains largely unchanged from the case of constant vorticity, where the free-surface overhangs and nearly self-intersects. In this sense, the constant vorticity upper layer seems largely unaffected by a change in the flow behaviour of the lower layer. 

\section{Conclusion and future work}

We have presented a numerical discretisation of the formulation introduced and analytically examined by \cite{wahlen2023global} for periodic travelling water waves under gravity for an arbitrary distribution. Employing a conformal mapping rather than a semi-hodograph transformation, our approach allows for internal stagnation points, critical layers, and overhanging profiles, broadening the range of computable solutions compared to previous studies \citep{dalrymple1977numerical,thomas1990wave,ko2008large,ko2008effect,amann2018numerical}. We have tested our numerical scheme against previously computed solutions for zero and constant vorticity, finding strong agreement. 
 
We have uncovered novel solutions for affine vorticity functions, revealing intricate internal flow configurations, including multiple critical layers, theoretically predicted by \citet{ehrnstrom2011steady,ehrnstrom2012steady}. Furthermore, we have computed solutions where a smooth vorticity transition separates two layers of constant vorticity. When a sign change occurs in the vorticity, we observe that large-amplitude solutions develop near-stagnant regions in the transitional layer.

Some drawbacks of our numerical approach have been identified. While conformal mapping from a rectangular auxiliary domain allows for internal stagnation points, it does not permit singularities at the boundary. This presents challenges when a solution branch tends to stagnation at the fluid surface, because the local flow behaves like that inside a $120^\circ$ corner, thereby the mapping is singular there. Additionally, since the method requires solving the problem throughout the fluid domain, rather than only at the fluid surface, as is for zero and constant vorticity, it can become computationally impractical to resolve some large-amplitude solutions.

Future work will involve a more comprehensive exploration of the solution space, focusing on both physically relevant and mathematically interesting vorticity functions. Additionally, modifications to the numerical method could allow for singular vorticity distributions in the bulk of the fluid and singularities at the boundaries, utilising a function splitting method developed by \cite{woods1953relaxation}. Finally, an intriguing potential extension is to consider vorticity induced by density variations, that is, baroclinic vorticity, where the density is given as a function of the stream function. See, for instance, \cite{long1953some}. The methodology developed herein can be adapted to study free-surface solitary waves in a stratified fluid, replacing the rigid-wall assumption commonly used in previous studies 
\citep[see][and references therein]{stastna2022internal}.

\section*{Acknowledgements}
AD would like to acknowledge funding from EPSRC NFFDy Fellowships (EPSRC grants EP/X028607/1). VMH acknowledges funding from US NSF DMS-2407358.

\section*{Data availability}
 The codes used to produce the solutions seen in this paper (written in MATLAB), are available at 

\noindent https://github.com/alexdoak. Solutions can be provided upon request of the corresponding author.

\bibliography{main}  
\bibliographystyle{abbrvnat}

\end{document}